\documentclass{eas}
\usepackage{graphicx}
\begin{document}

\TitreGlobal{Mass Profiles and Shapes of Cosmological Structures}

\title{Unifying dark energy and dark matter with a scalar field}
\author{Arbey, A.}\address{Centre de Recherche Astronomique de Lyon (CRAL), 
9 Avenue Charles Andr\'e, 69561 Saint Genis Laval Cedex, France -- Email: arbey@obs.univ-lyon1.fr}
\runningtitle{Unifying dark energy and dark matter with a dark fluid}
\setcounter{page}{1}
\index{Arbey, A.}

%
\begin{abstract}
The standard model of cosmology considers the existence of two components of unknown nature, ``dark matter'' and ``dark energy'',  which determine the cosmological evolution. Their nature remains unknown, and other models can also be considered. In particular, it may be possible to reinterpret the recent cosmological observations so that the Universe does not contain two fluids of unknown natures, but only one fluid with particular properties. After a brief review of constraints on this unifying ``dark fluid'', we will discuss a specific model of dark fluid based on a complex scalar fluid.
\end{abstract}

\maketitle

%
\section{Introduction}
In the standard model of cosmology, the total energy density of the Universe is dominated today by the densities of two components: the ``dark matter'' has an attractive gravitational effect like usual matter, while the ``dark energy'' can be considered as a kind of vacuum energy with a negative pressure, which seems constant today. The real nature of these two components remains unknown, but dark matter is generally modeled as a system of collisionless particles, and Usual models for dark energy are the quintessence models, based on scalar fields. However, many problems still remains in usual dark energy and dark matter models, and other models are worth to be investigated. In this paper, I will consider a model in which the dark matter and the dark energy are in fact different aspects of a same fluid, the ``dark fluid''. I will first review several observational constraints and I will provide ideas on how to build a dark fluid model thanks to a complex scalar field.

\section{Observational constraints on the dark fluid}
Many cosmological observations can be used for defining constraints on the dark fluid explaining at the same time the dark matter and the dark energy problems. A deeper analysis of the constraints was performed in (Arbey, 2005). I will review here some results from this study.\\
Let us define $\Omega_D$ as a ratio of the density of the dark fluid over the critical density, and $\omega_D$ as a ratio of the pressure of the dark fluid over its density. The observations of supernov\ae~of type Ia lead to the following constraints on the dark fluid:
\begin{equation}
\nonumber \Omega_D^0 = 0.962 \pm 0.084 \;\;\;\;\;\;\;\omega_D^0 = -0.76 \pm 0.25 \;\;\;\;\;\;\;\omega_D^1 =  1.0 \pm 0.6 \;\;,
\end{equation}
where $\omega_D$ is approximated, at low redshift, as $\omega_D=\omega_D^0+ \omega_D^1 z$.\\
From the structure formation, no stringent constraints can be determined without performing a precise analysis of the specific dark fluid model. A quick analysis nevertheless shows that $\omega_D > -1/3$ at the time of structure formation.\\
Considering the Cosmic Microwave Background, one can show that a large variety of dark fluid models is permitted. One can however consider that a model is potentially a ``good'' dark fluid model if the fluid behaves today like a cosmological constant whereas it could have behaved mainly like matter at last scattering.\\
At the time of the primordial nucleosynthesis (BBN), if it is correct to consider a Universe dominated by radiation, the main constraint is that the dark fluid density should be small in comparison to the radiation density; it means that, if one assumes that the dark fluid behavior does not change violently during BBN, the equation of state of this fluid around the time of BBN has to be $\omega_D(\mbox{BBN}) \leq 1/3$, or that its density was completely negligible before BBN. In the case of a real radiative behavior $\omega_D(\mbox{BBN}) = 1/3$, the dark fluid behaves like extra--families of neutrinos, and its density can be constrained.\\
These constraints can seem not very stringent if one considers them separately, but a complete dark fluid model would have to agree with all of them, which can seem difficult to achieve. We will now consider whether a scalar field can be used to build a dark fluid model.

\section{A complex scalar field as dark fluid}
\begin{figure}[h]
   \centering
   \includegraphics[width=4.3cm,angle=270]{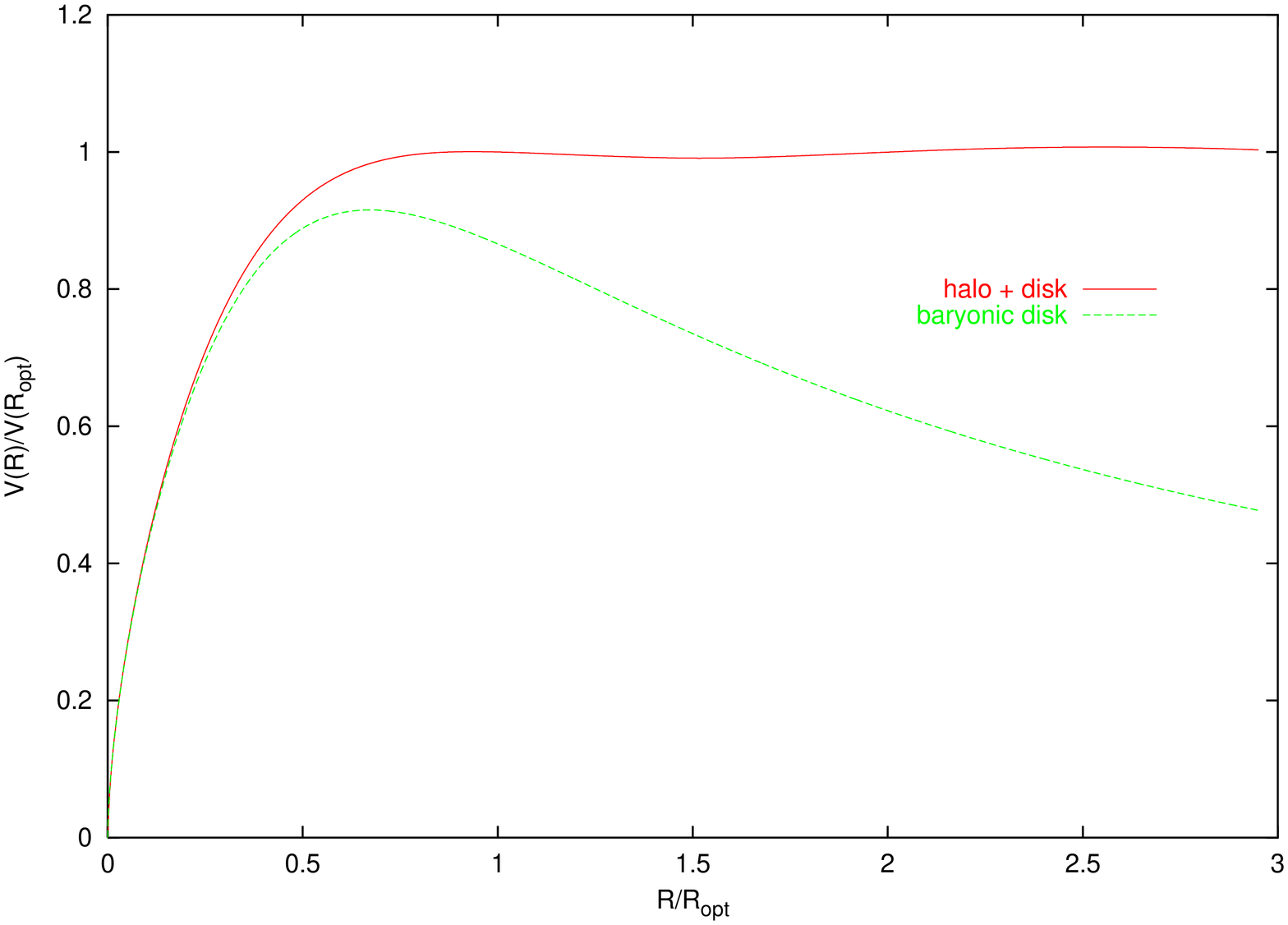} \includegraphics[width=4.3cm,angle=270]{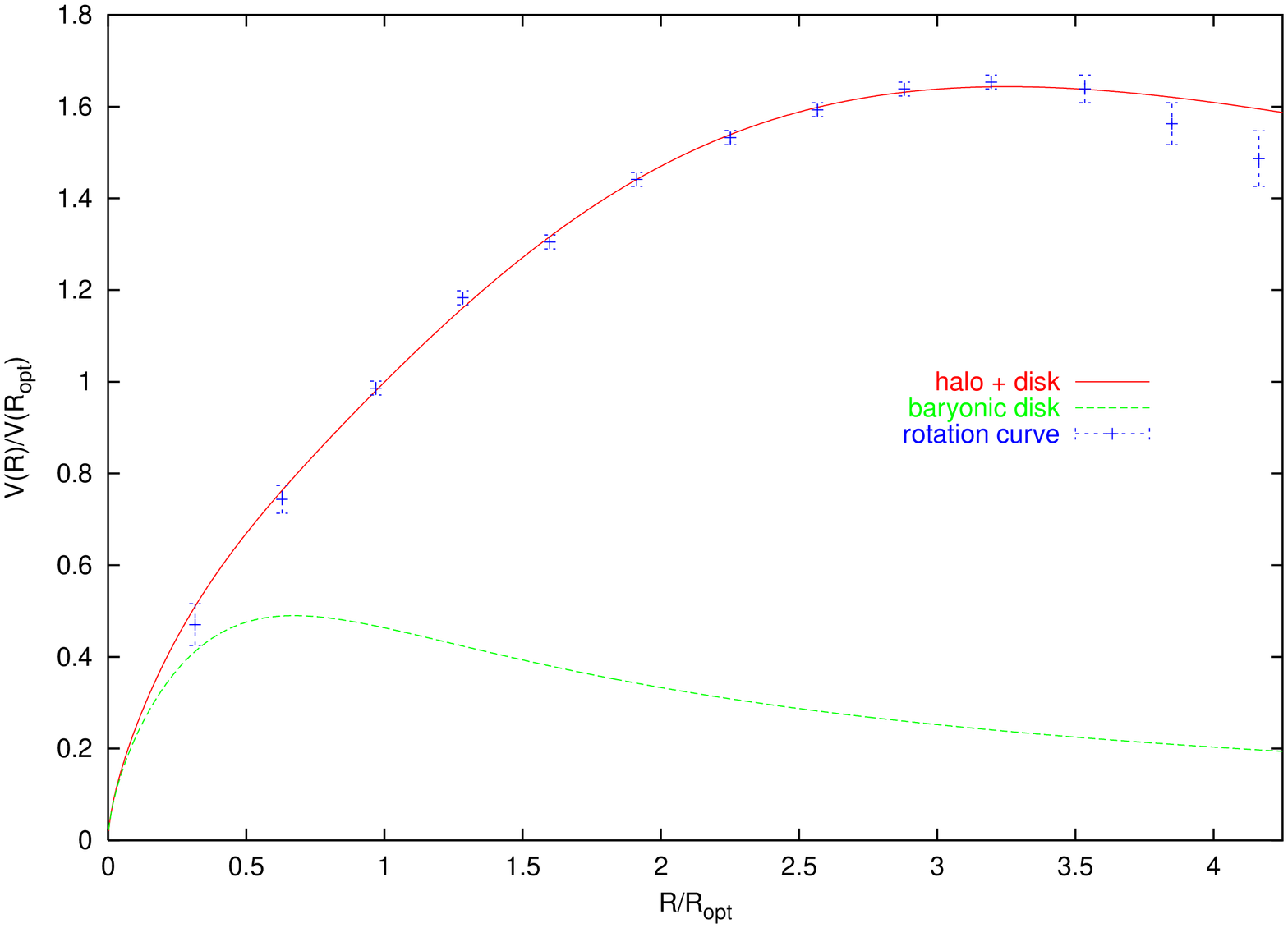}\
   \caption{On the left, flat galaxy rotation curve (solid line) induced by the presence of a complex scalar field with a quadratic potential. The dashed line shows the contribution from the baryonic disk only. On the right, rotation curve of the dwarf spiral DDO 154 fitted with the same field.}
       \label{gal}
   \end{figure}

\begin{figure}[h]
   \centering
   \includegraphics[width=4.4cm,angle=270]{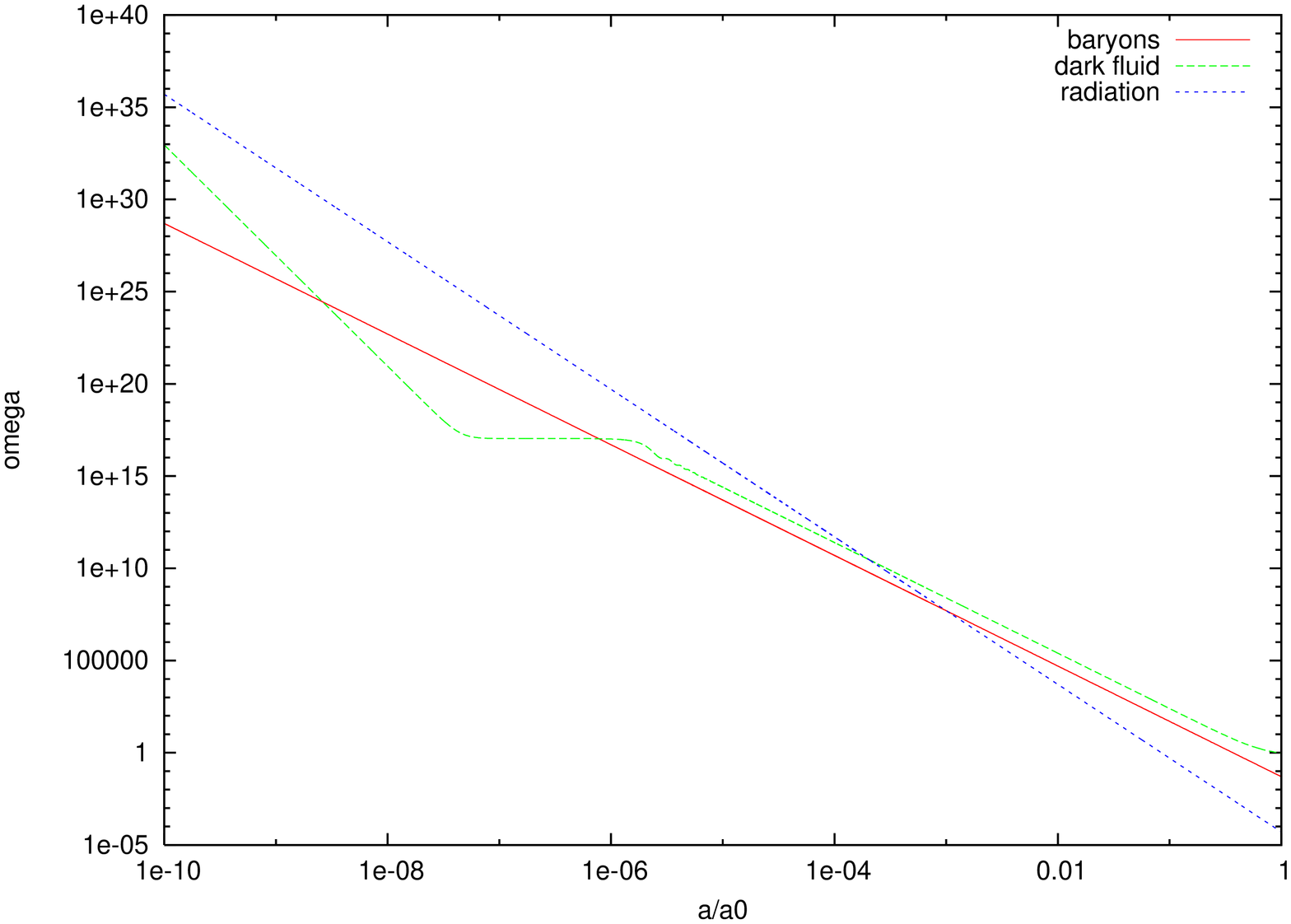}\includegraphics[width=4.4cm,angle=270]{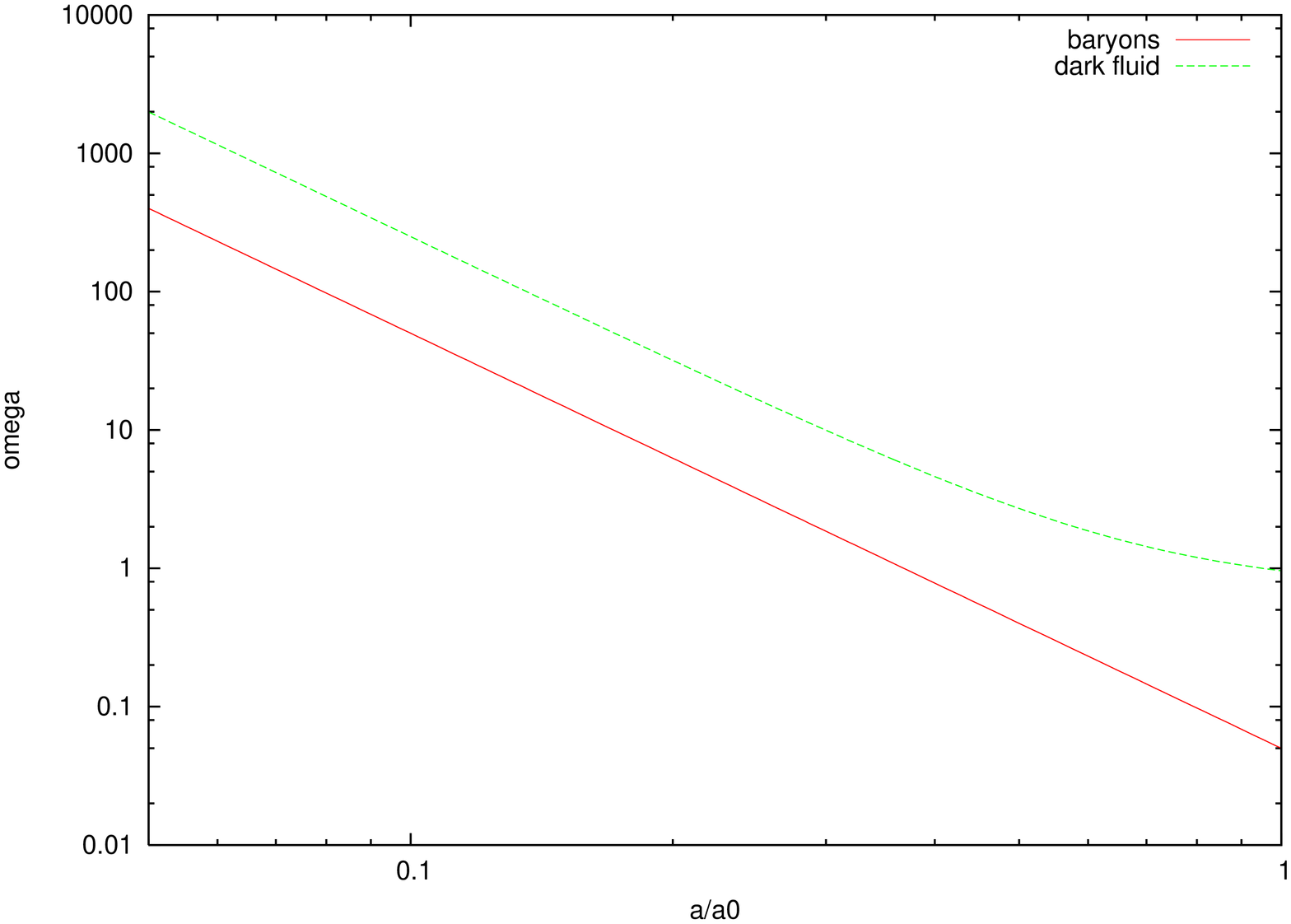}
   \caption{Cosmological evolution of the density of a complex scalar field in comparison to the densities of baryonic matter and radiation.}
   \label{cosmo}
   \end{figure}

\noindent In the literature, only several fluids behaving like a dark fluid are considered, and the Chaplygin Gas is the most known of them (see Kamenshchik et al., 2001, Bilic et al., 2002 and R. Bean et al., 2003). In this article, I will consider in particular a model using scalar fields.\\
Scalar fields have been often used to explain the dark energy problem, and refered as quintessence scalar fields  (Peebles et al., 1988). The main parameter for these models arises from the potential of the scalar field, and one can note for example that exponential potentials have been highly investigated (and excluded by supernov\ae~observations). Many other potentials are also under investigation.\\
However, a dark fluid model can be built thanks to a scalar field only if this field can also behave like dark matter. A previous study has shown that a complex scalar field can account for dark matter in galaxies and in cosmological evolution (Arbey et al., 2001/2002/2003). We have shown in particular that if one considers a complex scalar field $\varphi$ with an internal rotation $\varphi = \sigma e^{i\omega t}$ and based on a lagrangian density:
\begin{equation}
{\cal L} \; = \;g^{\mu \nu} \, \partial_{\mu} \varphi^{\dagger} \, \partial_{\nu} \varphi\; - \; V \left( |\varphi| \right) \;\; ,
\end{equation}
associated to a quadratic potential $V(\varphi) =m^2 \varphi^\dagger \varphi$, it is possible to reproduce rotation curves of spiral galaxies (see for example fig.~\ref{gal}). So, one can hope to bluid a dark fluid model from a complex scalar field whose potential contains a $m^2 \varphi^\dagger \varphi$ term.\\
Therefore, it is known that scalar fields can account either for dark matter or for dark energy. In both cases, the behaviour of the scalar field is determined by its potential. If one wants to use the same scalar field to unify both problems, the greatest difficulty is to determine which potential can provide a matter behaviour at local scales and satisfy the cosmological constraints described in the precedent section. A good candidate model would consist in using a complex scalar field associated to a potential which is a superposition of a quintessence potential and of a quadratic potential. For example, one can investigate:
\begin{equation}
V(\varphi) =m^2 \varphi^\dagger \varphi + A \exp(-B |\varphi|) \;\;.
\end{equation}
This potential can lead to the cosmological behaviour of figure \ref{cosmo} provided the different parameters are chosen adequately. 
These parameters have to be chosen so that until structure formation, the quadratic part of the potential dominates and the field behaves like matter, and very recently the second part of the potential dominates, leading to a quintessence behaviour. The case described on the figure is consistent with the constraints previously presented. An important question is now to know if the behaviour of the fluid can be matter-like at local scales and quintessence-like at cosmological cases. To answer briefly to this question, one can remark that the density of dark fluid on cosmological scales today is of the order of the critical density, i.e. $\rho^0_c\approx~9\times~10^{-29}\mbox{g.cm}^{-3}$, whereas the estimated matter density in the Milky Way at the radius of the Sun is $\rho^{\mbox{\tiny Sun}} \approx 5 \times 10^{-24}\mbox{g.cm}^{-3}$. Hence, even if the local density of the dark fluid would represent only 1\% of this total local matter density, its value would be much higher than the cosmological densities today. It means that, where the density of scalar field is high, for example around galaxies or in the Early Universe, the quadratic part of the potential has to dominate, and where the density is low, the second part of the potential dominates, leading to a repulsive vacuum energy-like behaviour. Then, such a dark fluid model leads to a Universe highly inhomogeneous today.
   
\section{Conclusion}
\noindent Astrophysical and cosmological observations are usually interpreted in terms of dark matter and dark energy. We have seen here that they can also be analyzed differently. Thus, it is possible to develop a model of dark fluid which could advantageously replace a model containing two dark components. The properties of the dark fluid are different from dark matter and dark energy, and are worth to be studied. A more complete study of the potentiality of complex scalar fields for building dark fluid models will be provided elsewhere.


\end{document}